\input{aipcheck}

\documentclass[final]{aipproc}

\layoutstyle{6x9}

\begin{document}

\title{PULSAR RADIO EMISSION CUTOFFS }

\classification{ \texttt{97.60.Gb, 94.30.Gm, 95.30.Gv}}
 
\author{Ericsson D. L\'{o}pez}{
  address={Observatorio Astron\'{o}mico de Quito
\\ Interior del Parque La Alameda, (17-01-165) Quito, Ecuador\\}
}

\begin{abstract}
The propagation of radio emission in pulsar  magnetospheres is discussed.  We follow a kinematic model in order to derive dispersion relations for electromagnetic oscillations and transversal waves, propagating in a cold moving plasma. We have included relativistic corrections on the dispersion properties, involved with the relativistic motion of the emitting plasma. The occurrence of  plasma instabilities is analyzed  beside the conditions which should be fulfilled in order to permit the wave propagation and conversion in regions close to the cutoffs of the system.   The existence of various frequencies of resonance  has been predicted and we are working out  these results in order to explain the low-frequency  cutoffs  observed in radio pulsar espectra.

\end{abstract}

\maketitle

\section{ Introduction }
Many problems in current astrophysics, plasma physics and magnetohydrodynamics are related to the theory of transport radiation through a moving medium. For example, the radiation originated in a neutron star must travel through the moving magnetosphere corotating with the central compact object. Moreover, assuming that the corotation velocity varies with the distance from core,  at certain distance from the core the moving medium becomes relativistic and one may expect anomalous behavior from the radiation passing through the plasma.

Recently, several papers have been devoted to study this kind of problem. For instance, \cite{lu01} pointed out in his  review, the processes according to which the magnetospheric plasma, in pulsars, affects propagation of radio emission. There are four main factors that   can influence the propagation: reflection, refraction, scattering and absorption. Regarding the absorption, the waves can be absorbed by resonance and scatter in processes involving interaction with plasma particles or another wave.  Furthermore, it is supossed that the radio emission is generated in the pulsar outflowing plasma \citep{me00} and the propagation effects take place inside the light cylinder. Three main resonances processes have been identified: the Cherenkov resonance \citep[e.g.][]{ar86}, Cherenkov-drift resonance \citep[e.g.][]{ly99} and the cyclotron resonance \citep[e.g.][]{mi79,ly99}.  In order to explain the features of the radio emission observed in pulsars, the above  mechanisms beside those corresponding to the transverse electromagnetic waves generation, have been considered.  Many papers have been devoted to discuss  the kind of electromagnetic waves which are able to propagate in a pulsar plasma, however, the pulsar radio emission models up to now proposed are not able to explain all the main features of observational facts. Futhermore efforts are neccesary in order to incorporate, for instance, relativistic effects which may conduct to more detailed descriptions of the absoption and resonance instabilities, which are supossed to produce electromagnetic waves and oscillations in pulsar plasmas.

    In this paper we  follow the results of our previous dispersion model \citep{lo04} developed for relativistic moving magnetoactive plasmas, in order to establish the electromagnetic spectra and the modes  corresponding to absorption cutoffs and instabilities, which are propagating in the  pulsar magnetosphere as well as in the wind driven by the open magnetic field lines at the polar caps.  We show, that the relativistic effect on the dispersion properties of a magnetoactive medium appears to be important when the motion of the plasma occurs with highly relativistic velocities and  the girofrequency $w_b$ is of the order of plasma frequency $w_p$. The combined action of the magnetic field and bulk velocity increases the influence
of the medium on the dispersion processes, which are dependent on the direction of the wave propagation.

 In  section 1, we describe briefly the dispersion model that we  have used and present the plasma solutions that are able to propagate in the pulsar plasma. In  section 2, the observational features of the radio emission in pulsar are stated and  we interpretate the observations according with the results of the proposed dispersion model.

\section{ The Relativistic Treatment of an Anisotropic Magnetoactive Plasma }

The influence of the global plasma motion on the dispersive properties of the medium has been
recently studied by \citet[][]{lo04,lo96a,lo96b}.
Classical treatment of a magnetoactive plasma shows that the global motion does not affect the dispersion properties of the medium. The normal waves should propagate like those in a non-moving system. However, if the plasma is moving at relativistic velocities, its influence on eventual anisotropies turn out to be more substantial and new relativistic effects appear. For instance, the propagation of new electromagnetic modes becomes possible besides the ordinary waves of a magnetoactive plasma \citep[][]{lo96a, lo96b}.

We consider a plasma of low temperature consisting only of electrons and ions, and let both components  move relative to each other in equilibrium. Then the low temperature of the plasma makes it possible  to treat the particles classically (${v_t\over c} \ll 1$, where $v_t$ is the thermal velocity). We consider, in addition, the collisionless regime.
All of our calculations are translated to the laboratory frame. We begin by finding the suitable dielectric tensor of permittivity $\epsilon_{ij}$ for a magnetoactive plasma which is moving with respect to
the laboratory frame. Using the Lorentz's transformation, a tensorial relation between the electric vector $\vec E'$ in the rest frame and the vector $\vec E$ in the laboratory frame can be achieved. A
similar relation for the electron currents $\vec j$  and $\vec j' $  is also settled. With this, we may
derive a tensorial relation for the transformation of the tensor of the conductivity $\sigma_{ij}$,
which is related with the dielectric tensor $\epsilon_{ij}$ by the well known expression \citep{gi60}:

$$\epsilon_{ij}(w,\vec k) = \delta_{ij} + {{4 \pi i} \over w} \sigma_{ij} ( w, \vec k).$$

\noindent
Following this procedure it is not difficult to derive the relativistic
relationship between the elements of the permittivity tensor
in different coordinate systems. It is the relativistic generalization of the
well known Minkowsky's relation and it is given by \citep{lo96a}:

        $$\epsilon_{ij}(w,\vec k)=\delta_{ij} (1- { w'^2 \over {w^2}})+{w'^2 \over
w^2}\epsilon'_{ij}- {w'\over{w}}\gamma(1-\beta){{\epsilon'_{i\nu}v_\nu v_j}\over {v^2}}
{w'\over{w^2}}\gamma \epsilon'_{i\nu}k_\nu v_j + $$ $${w'\over{w}}\gamma[(1-\beta){{v_iv_j}\over{v^2}} -
{{v_jk_i}\over{w}}]+ 
{w'^2\over{w^2}}{{\gamma(1-\beta)}\over {v^2}}(v_i\epsilon'_{\mu j}v_{\mu}-v_iv_j)
{w'\over{w}}{\gamma^2\over {v^4}}(1-\beta)^2 v_iv_j $$ 
$$(v^2-v_\mu \epsilon'_{\mu \nu}v_\nu)+
 {w'\over{w^2}}{{\gamma^2(1-\beta)}\over {v^2}} v_iv_j(v_\mu \epsilon'_{\mu \nu}k_\nu-v_\mu k_\mu)+ 
{w'\over{w^2}}\gamma v_i(k'_\mu \epsilon'_{\mu j}-k'_j)- $$
$${{\gamma^2 v_iv_j}\over {w v^2}}(1-\beta)
 (k'_\mu\epsilon'_{\mu \nu}v_\nu -v_\mu k'_\mu) + {\gamma^2\over {w^2}}v_iv_j
 ( k'_\mu\epsilon'_{\mu \nu}k_\nu - k'_\mu k_\mu), \eqno(1) $$

\vskip 0.3cm
\noindent
where $\epsilon_{ij}(w,\vec  k)  $  is  the  dielectric  permittivity
tensor  in  the laboratory frame $xyz$,~  $\epsilon'_{ij}(w',\vec k')$
is the dielectric permittivity tensor in the rest frame $x'y'z'$ of the plasma,~ $w$ is
the  radiation  frequency in the system $xyz$,~ $w'$ is the radiation frequency
in $x'y'z'$ system,~  $\vec  k$  and  $\vec  k'$ are the wave vector in the $xyz$ and $x'y'z'$
systems, respectively,~ $ v(x,y,z)$ is the plasma velocity, and:

$$ \gamma = {1\over{\sqrt{1-{v^2\over c^2}}}} \hskip 1cm \beta = {1\over{\gamma}}$$
$$ w' = \gamma(w-\vec k . \vec v) \hskip 1cm \vec k' = \vec k + {\gamma\over {v^2}}
[(1-\beta) \vec k.\vec v - w{{v^2}\over {c^2}}] \vec v. $$

For the description below, the adopted coordinate system is such that the plasma is flowing along the z-axis parallel to the magnetic
field lines, the so-called longitudinal flow, which corresponds to a situation typically
observed in astrophysical outflows. Then, the global plasma velocity
will be $\vec v = v_z \vec k$, and the tensor above can be written in the simple matrix form as:

$$ \epsilon = I + {w'\over w}\alpha (\epsilon'-I)\beta \eqno(2) $$

\noindent
where I is the unitary matrix, $\epsilon'$ is the dielectrical permittivity matrix in the
co-moving system and

                 $$\alpha=\pmatrix { 1 &0& 0\cr
                                     0&1&0\cr
                         {\gamma v k'_x}\over {w'}&0&1+
                         \gamma [(1-\beta)+{{vk'_z}\over{w'}}]\cr},  \hskip 1cm
\beta=\pmatrix { {w'}\over{w} &0& {{\gamma v k'_x}\over {w}}\cr
                                     0&{w'\over w}&0\cr
                         0&0&{w'\over w}-\gamma [(1-\beta)-{{vk_z}\over{w}}]\cr}.$$

\vskip 0.3cm
The tensor $\epsilon'_{ij}$ for a magnetoactive plasma has the usual form:

    $$\epsilon'=\pmatrix { \epsilon_\perp &ig& 0\cr
                            -ig&\epsilon_\perp&0\cr
                         0&0&\epsilon_\parallel\cr},\eqno(3)$$
\noindent
where:

$$\epsilon_\perp=1- { w_p^2\over { w^2-w_b^2 }}
\hskip 1cm
g={-w_p^2w_b\over{w(w^2-w_b^2)}}
\hskip 1cm
\epsilon_\parallel=1- { w_p^2\over w^2}$$

\noindent
and $w_p=5.64x10^4 \sqrt{n_e}~ $ Hz , $w_b=1.76x10^7 ~ ({B\over \gamma}) $  Hz  are the plasma frequency and
the gyrofrequency, respectively, $n_e$ is the electron density in $cm^{-3}$, B is the magnetic field
in $G$ and $\gamma$ is the Lorentz factor.

Substituting Eq.(3) into Eq.(2), the dielectrical permittivity tensor for a relativistic
moving magnetoactive plasma is derived. This tensor looks like a hermitician matrix for this particular jet configuration, its components are presented here:

     $$\epsilon =\pmatrix { \epsilon_{xx} & \epsilon_{xy}&\epsilon_{xz} \cr
                                    -\epsilon_{xy}&\epsilon_{xx}&\epsilon_{yz}\cr
                         \epsilon_{xz}&-\epsilon_{yz}&\epsilon_{zz}\cr},\eqno(4)$$
\noindent
with:

$$ \epsilon_{xx} = - {{c^2k^2v^2(w_p^2-w^2)cos^2\theta + 2c^2 k v w(w^2-w_p^2) cos\theta
}\over
{w^2[c^2k^2v^2 cos^2\theta- 2c^2 k v w cos\theta + c^2(w^2-w_b^2)+w_b^2v^2]}}$$
$$
+ {{w^2[c^2(w_b^2+w_p^2-w^2)-w_b^2v^2]}\over
{w^2[c^2k^2v^2 cos^2\theta- 2c^2 k v w cos\theta + c^2(w^2-w_b^2)+w_b^2v^2]}}$$
$$ \epsilon_{zz} =1+ {{w_p^2v^2}\over{c^2(kv cos\theta-w)^2}}-
  {{w_p^2}\over {(kv cos\theta -w)^2}}-$$
  $$ {{c^2k^2w_p^2v^2 sin^2\theta}\over
{w^2[c^2k^2v^2 cos^2\theta-2 c^2 k v w cos\theta + c^2(w^2-w_b^2)+w_b^2v^2]}} $$

$$ \epsilon_{xy} = {{icw_bw_p^2 \sqrt{c^2-v^2} (kv cos\theta - w)}\over
{w^2[c^2k^2v^2 cos^2\theta-2c^2k v w cos\theta + c^2(w^2-w_b^2)+w_b^2v^2]}}$$

  $$ \epsilon_{xz} = {{c^2kw_p^2 v sin\theta (kv cos\theta -w)}\over
{w^2[c^2k^2v^2 cos^2\theta-2c^2 k v w cos\theta + c^2(w^2-w_b^2)+w_b^2v^2]}}$$

$$ \epsilon_{yz} = {{ickw_bw_p^2v \sqrt{c^2-v^2} sin\theta }\over
{w^2[c^2k^2v^2 cos^2\theta- 2 c^2 k v w cos\theta + c^2(w^2-w_b^2)+w_b^2v^2]}}$$

Using the Maxwell's equations together with the material equation $D_i = \epsilon_{ij}~ E_j$, and adopting
the assumption that the perturbation has the form $ exp~ i( \vec k.\vec r - w t)$, the algebraic equations that describe the wave dispersion in a plasma can be derived and expressed as:

$$ [k^2 \delta_{ij} - k_i k_j - {w^2\over c^2} \epsilon_{ij}(w,\vec k)] \vec E_j = 0 $$

The condition to solve this system of equations defines the electromagnetic waves which are allowed to propagate inside a moving magnetoactive
plasma. This condition, with the application of some determinant properties, may easily be converted into the following
tensorial dispersion equation \citep[see][]{lo04,lo96a,lo96b}:

    $$ {{k_ik_j\epsilon_{ij}}\over k^2} k^4 + [{{k_ik_j}\over k^2}\epsilon_{i\mu}\epsilon_{\mu j}
     -\epsilon_{ii}{{k_ik_j\epsilon_{ij}}\over k^2}] [{w^2\over c^2} k^2]
      + \mid \epsilon_{ij}\mid {w^4\over c^4} = 0. \eqno (5)$$

Now, we must use  Eq.(5) along  with the dielectric tensor of permittivity given by Eq.(4), in order to derive the electromagnetic spectrum characteristic of a moving magnetoactive plasma. The dispersion Eq.(5) may be solved for an arbitrary dispersion angle $\psi$, that is, the angle between the jet velocity and the direction of the radiation propagation defined by the wave vector  $\vec k $ \citep{lo96a,lo96b}. This is a large and complex tenth order equation in $w$ containing 123 terms, which does not admit exact solutions except for some particular field-velocity orientations.

 The above formulation can be applied to any  magnetoactive plasma which is  moving parallel to the magnetic field. For these systems, we will notice that the higher the relativistic velocity of the plasma the more important the relativistic effects over the dispersive properties are.

In what follows, we discusse the  longitudinal modes, these ones have refraction indices $n \gg 1$ and  excite pressure waves and turbulence in the plasma.The other cutoffs are contained in the  transverse electromagnetic normal modes,  disussed in a separated paper \citep[e.g.][]{lo04}.

\subsection{ Longitudinal wave propagation }

 In this case we considere that the electromagnetic radiation is propagating along the magnetic field, then the permissible transverse electromagnetic modes that could propagate in a magnetoactive moving plasma are described by the following dispersion relation, which is derived analytically \citep[e.g.][]{lo04}:

$$n_{1,2}^2 = 1\pm {{w_p^2[{w_b\over \gamma}\mp (1-{v\over c}) w]}\over
{w[(1-{v\over c}) w^2 - (1 + {v\over c}) w_b^2]}} \eqno(6) $$

The corresponding curves are similar to the dispersion curves for waves propagating along the magnetic field lines in a static magnetoactive plasma. But in a moving plasma the  extraordinary branch, have a cutoff near the gyrofrequency ($w_b$) with a strong dependence on the plasma velocity:

$$ w_r = (1+{v \over c})~ \gamma~ w_b .\eqno(7) $$

\noindent
For small velocities ${v^2\over c^2} \ll 1$, the above expression falls into the classical ordinary and extraordinary solutions for a non-moving plasma \citep{gi60}:

$$ n_1^2 = 1 - {w_p^2\over{w(w+w_b)}}   \hskip 1.5cm   n_2^2 = 1 - {w_p^2\over{w(w-w_b)}}.\eqno(8)$$

\noindent
The other analytical solution of the dispersion equation corresponds to the longitudinal waves which are described by:

$$ n_{3,4} = {c\over v}\big[1 \pm {w_p\over {w}}\sqrt{(1-{v^2\over c^2})} \big] \eqno(9)$$

Therefore, for the radiation propagation along the magnetic field, we have identified two cutoffs: one at  the  frequency ${w_b\over \gamma}$ and the second given by the Cherenkov-like resonance $w-\vec k . \vec v - {w_p\over \gamma} = 0$. At this point, it is worthly to note that the latest Doppler-shifted resonance appears as if the   magnetic field is not present. This is in contrast, to that stated in the literature, where a hybrid Cherenkov-cyclotron resonance has been derived from a magnetohydrodynamic point of view \citep[][]{ly99,lou02}.  

\subsection{ Transverse wave propagation}

For the case when the wave is propagating perpendicular to the direction of
the magnetic field ($\psi = 90^{0} $), the Doppler effect becomes quadratic
and the radiation is characterized by the following analytic dispersion relations:

$$ n_{1,2}^2 = (1- {w_p^2\over w^2})\pm {{{w_b\over w} {w_p^2\over w^2}[\sqrt{{w_b^2\over w^2} +
4{v^2\over c^2}(1- {w_b^2\over w^2}- {w_p^2\over w^2} )}-{w_b\over w} ]}
\over{2[(1- {w_b^2\over w^2}- {w_p^2\over w^2} ) +
{v^2\over c^2}({w_b^2\over w^2}+{w_p^2\over w^2})]}}.  \eqno(10) $$

 In this case, the frequencies for the points of resonance are given by:

$$ w_r = 0,  \hskip 2cm
 w_r = {\sqrt{(w_{b}^{2} + w_{p}^{2})}\over \gamma }.\eqno(11)$$

\vskip 0.5cm
Also here, we touch only the existence of vacuum-like resonance, i.e. electromagnetic waves that may leave freely the magnetized plasma. Furthermore, as we must expect, in the transverse direction we do not have any longitudinal oscillation \citep{lo96a,lo96b}. Consequetly,  in this direction we do not have instabilities due to the electron-wave interaction.
The analitical  solutions from the general dispersion equation (4), produce again two
different electromagnetic branches. One of these solutions corresponds to an ordinary
wave ($n_1^2$) and the other one to an extraordinary wave ($n_2^2$). 

Under the assumption that the outflow moves slowly (classical aproach), we have different points of oscillation that may develop instabilities. For instance, considering that the cold magnetoactive plasma is moving along the magnetic field, we have found that the Cherenkov-like plasma oscillations are  splitted  into the cyclotron-like and plasma frequency-like, i.e., these instabilities are desconnected:

 $$ w - \vec k . \vec v  - w_b = 0 ~~~~~~~~w - \vec k . \vec v - {w_p} = 0.\eqno(12)$$

On the other hand,  if the plasma is moving transverse to the magnetic field lines, the cyclotron and plasma instabilities appear connected and cyclotron plasma Cherenkov-like oscillations are developed:

$$w - \vec k . \vec v - w_p - w_b = 0\eqno(13)$$.

Furthermore, it is feasible to show that  Cherenkov-like instabilities also appear for an arbitrary direction of the waves. However, in the general case,  the related expressions are more complex and the dispersion relations and cutoffs should be obtained applying numerical methods. 

\section{Pulsar radio emission}

Owing  to energetic charged particles  which are produced inside the light cylinder, close to the core, the pulsars are characterized by a radiating magnetosphere. The plasma instabilities seem to be responsable for developing  relativistic particle beam, propagating along the open field lines to merge into powerful pulsar wind \citep[e.g.][]{go70}.  The wind is moving relativistically at the light cylinder region, whereas the corotating magnetosphere is also moving with   highly velocities.  Then, a dispersion model taking into account the bulk motion of these media could be able to suggest how the radiation, that is passing through, is affected. 

On the other hand, the features of normal modes  also should be predicted by such a dispersion model.  Therefore, the low-frequency cutoffs observed between 100 MHz  and 1 GHz in radio pulsa espectra \citep[e.g.][]{ma94,me99}, may be physically explained.
The ocurrence of low-frequency cutoffs observed in the profiles of the pulsar emission are believed to take place in the magnetized plasma of the pulsar magnetosphere. Equations  (7) and (11) predict the frequency cutoffs for radio emission coming from these regions. In both expressions, the cyclotron $w_b$ and plasma $w_p$ frequencies (natural plasma frequencies), have been attenuated by the plasma bulk $\gamma$-factor.  We note also that the apropiate cutoff will take place in dependence of  magnetic field orientation. For the inner regions toward the core, the cyclotron resonance is most probable (see Eq. 7), whereas beyond the light cylinder the hybrid $w_b$ and $w_p$ resonace, given by the equation (11), is more feasible. For estimates, we assume the typical plasma parameters at the emission region, $R\sim 10^9~cm$ \citep{ly99}. At the plasma rest frame the cyclotron frequency $w_b =1.8 \times 10^{10}~rad~s^{-1}$ and the plasma frequency $w_p = 1.2\times 10^{8}~rad~s^{-1}$. Then, the corresponding frequency cutoff attenuated by plasma bulk $\gamma= 10 $ should be $\nu_r \sim 0.3~GHz$, value  of the order of the typical frequency of emission, $\sim 1~GHz$, of pulsars \citep{me99}. For the inner regions at the surface of the neutron star, the equation (7) gives us a much higher frequency cutoff $\nu_r \sim 2.9\times 10^{17}~ Hz$; assuming a dipolar magnetic field of $10^{12}~G$. Therefrom, we conclude that the observed radiation probably  is generated in  external emitting regions at  $R\sim 10^9~cm$ from the core, in agreement with previous works \citep[e.g.][]{ly99}.

On the other hand, the equations (9), (12) and (13) give us the existence in the moving plasma of  Cherenkov-like oscillations. Physically, these instabilities represent Doppler-shifted  resonances and generally reveal flow adventions. However, these plasma instabilities  are able to develop radiation by  maser-type mechanics, producing induced electromagnetic radiation.  The Cherenkov-like instabilities typically occur in the outer magnetosphere and can be handled in order to explain some observational facts present in the pulsar radio profiles.

\begin{theacknowledgments}
The author gives many thanks to the Brazilian agencies CLAF and CNPq that supported partially this work and the hospitality of the Instituto Astron\^{o}mico e Geof\'{i}sico, IAG-USP, where part of this paper was discussed previously. 
\end{theacknowledgments}

\end{document}